\documentclass[10pt]{article}
\usepackage{graphicx,floatflt,amssymb,epsf}
\def\a{\alpha}

\begin{document}
\begin{flushright}
\texttt{hep-ph/0407314}\\
SINP/TNP/04-13\\
\end{flushright}

\vskip 50pt

\begin{center}
{\Large \bf A comparison of ultraviolet sensitivities in universal, \\
nonuniversal, and split extra dimensional models} \\
\vspace*{1cm}
\renewcommand{\thefootnote}{\fnsymbol{footnote}}
{\large {\sf Paramita Dey} and {\sf Gautam Bhattacharyya} 
}
\vskip 5pt
{ \em Saha Institute of Nuclear Physics, 1/AF Bidhan Nagar,
 Kolkata 700064, India}\\
\normalsize
\end{center}

\begin{abstract}
We discuss the origin of ultraviolet sensitivity in extra dimensional
theories, and compare and contrast the cutoff dependences in
universal, nonuniversal and split five dimensional models. While the
gauge bosons and scalars are in the five dimensional bulk in all
scenarios, the locations of the fermions are different in different
cases. In the universal model all fermions can travel in the bulk, in
the nonuniversal case they are all confined at the brane, while in the
split scenario some are in the bulk and some are in the brane. A
possible cure from such divergences is also discussed.

\vskip 5pt \noindent \texttt{PACS Nos:~11.10.Kk, 12.60.-i} \\
\texttt{Key Words:~ Extra dimension, Kaluza-Klein tower}
\end{abstract}

\renewcommand{\thesection}{\Roman{section}}
\setcounter{footnote}{0}
\renewcommand{\thefootnote}{\arabic{footnote}}

\section{Introduction}
TeV scale higher dimensional theories \cite{anto} have been investigated from
the perspectives of high energy experiments, phenomenology, string theory,
cosmology and astrophysics. From a four dimensional (4d) point of view, a
higher dimensional field appears as a tower of 4d Kaluza-Klein (KK) states
labeled by ($n$). The multiplicity of KK states render all higher dimensional
theories nonrenormalizable. These are all effective theories, parametrized by
two additional quantities: the radius of compactification ($R$) and the
ultraviolet (UV) cutoff scale $M_S = (n_S/R)$. Even within the context of such
an effective framework, it is important to ask what is the UV sensitivity of
such a theory, i.e. approximately up to what scale one can perform a
perturbative calculation.  Admittedly, TeV scale extra dimensional theories do
not solve the hierarchy problem in a strict sense. But if particles with
nonzero gauge quantum numbers have access to the extra dimension then
experimental bounds push the inverse radius to at least a few hundred GeV or
approximately a TeV. Such theories may constitute the basic building block for
relatively more realistic models. Instead of looking for such realistic
constructions, all we aim in this paper is to consider simple and analytically
tractable, nevertheless experimentally allowed, toy scenarios and perform an
illustrative analysis of how the multiplicity of equispaced KK states
contribute to the nonrenormalizibility of such theories.  Here we consider
three scenarios, described below, and compare their UV cutoff dependences with
respect to different processes. We restrict to only one extra dimension and
assume a $Z_2$ discrete symmetry, i.e. the extra dimension is $S^1/Z_2$.

{\bf 1. Universal Extra Dimension (UED):}~ All particles are allowed
to access the extra dimension.  Its implications to oblique
electroweak parameters \cite{acd}, flavor changing neutral current
processes \cite{buras1,buras2,debrupa}, $Z\to b\bar{b}$ decay
\cite{santa2}, and other phenomenological processes
\cite{acd,uedothers} have been studied. All one loop processes turn
out to be finite because of a cancellation between wave function
renormalizations and vertex corrections. There is no dependence on
$M_S$.

{\bf 2. Nonuniversal Extra Dimension (NUED):}~ Fermions are localized at the
4d brane, while all bosons reside in the bulk. This is motivated \cite{pq}
from a stringy perspective that chiral matters should be placed in the twisted
sector while non-chiral states can travel in the bulk.  Constraints from
electroweak observables were placed on this scenario in \cite{ew}. This has
also been studied in the context of $Z\to b\bar{b}$ decay and Kaon and $B$
meson mixings \cite{santa1}. In a previous publication \cite{pdgb}, we probed
the root cause of UV sensitivity in this scenario with respect to some
electroweak loop processes, especially $B_d \to l^+l^-$.

{\bf 3. Split:}~ All bosons are in the bulk, but fermions are treated
differently in the sense that some fermions are in the bulk but some are
confined to the brane. Placing the fermions at different locations helps to
induce flavor structure \cite{hmos,branco} by generating different Yukawa
suppression factors for different fermions\footnote{In Ref.~\cite{branco},
  splitting refers to the idea of placing different fermions in different
  places inside a thick brane.}. {\em We define our split scenario as the one
  which has the first two generation of fermions in the bulk and the third in
  the brane}. Roughly, this picture is motivated in $S^1/(Z_2\times Z'_2)$ GUT
scenarios \cite{orbigut} for keeping consistencies with proton stability and
$b$-$\tau$ unification.  Neither is the way we have defined our split scenario
the only consistent bottom-up approach, nor can we justify it to follow from
some more fundamental consideration. All we want to emphasize is that although
this scenario is less elegant and rather {\em ad hoc}, nevertheless it is
neither more nor less realistic than either UED or NUED scenario.  It is all
the more important to study its UV sensitivity, which is different from that
of UED or NUED, for reasons that will be clear as we go along. We consider it
to be an illustrative example, and its other variants are not expected to
provide additional insights, so we do not consider them.

The crucial issue that controls the UV sensitivities is the question of KK
momentum conservation. For compact direction, momentum becomes discrete
($n/R$) but still remains conserved\footnote{In the UED scenario, what is
actually conserved is the KK parity \cite{matchev}. As a consequence, there
can be mixing only among even states or only among odd states, and that too
only at the orbifold fixed points. We do not consider such mixings as their
effects for our processes would be tiny.}. In UED scenario it is always
conserved. In split and NUED scenarios, the {\em localization} of some or all
fermions at the brane causes KK number nonconservation for a brane-localized
interaction. The issue of KK number conservation or nonconservation is
intimately linked to the occurence of a single or multiple KK sum in a loop
intergral involving KK modes in the internal lines. This aspect constitutes
the prime criterion to judge whether the theory would be well behaved or UV
sensitive. An illustrative example is given below.

Consider a conventional bosonic 4d propagator and its KK-towering. The
modification is as follows:
\begin{eqnarray} 
\label{kkinfinity}
{(k_E^2+M^2)}^{-1} \longrightarrow
\sum_{n=-\infty}^{\infty} {(k_E^2+M^2+{n^2}/{R^2})}^{-1}
= (\pi R/k'_E) \coth (\pi R k'_E),
\end{eqnarray} 
where $k_E$ is an Euclidean four momentum and $k'_E=\sqrt{k_E^2+M^2}$. For
large argument coth function goes like unity. This means that a sum over KK
modes reduces the power of $k_E$ in the denominator. Therefore, if a 4d loop
diagram is log divergent in the UV limit, then KK towering for only one
propagator turns it linearly divergent. If two propagators are separately
KK-summed (i.e. $n_1$ and $n_2$ are independent), then a logarithmically
divergent diagram becomes quadratically divergent, and so on. On the other
hand, if $n_1$ and $n_2$ are not independent by virtue of KK number
conservation, then the divergence will be less than quadratic.  A word of
caution is in order in this context.  Although an infinite KK summation before
actually performing the 4d momentum ($k_E$) integration is analytically doable
providing intuitive insight into the increasing hardness of the divergence
caused by the towering, the actual numerical coefficient of the divergence
turns out to be physicswise incorrect if infinite summation is performed first
and the 4d integration second (this point is emphasized in a specific case in
footnote 3). Towards the end, we discuss a possible mechanism which can cure
such divergences. Incidentally, we do assume that gravitational interaction
becomes important at a scale which is considerably higher than $M_S$ so that
we can safely neglect their contribution.

\section{The split scenario}
In this section, we explicitly write down the KK expansions of the
higher dimensional gauge, scalar and fermion fields in the split
scenario, when viewed from 4d perspective.

1. The extra dimension ($y$) is compactified on a circle of radius $R
   = M_c^{-1}$ and $y$ is identified with $-y$, i.e., it corresponds to
   an orbifold $S^1/Z_2$.

2. The gauge bosons $A^M(x,y)$, a generic scalar doublet $\phi(x,y)$
   and the first two generations of quarks and leptons are 5d
   fields. The 5d fields can be Fourier expanded in terms of 4d KK
   fields as:
\begin{eqnarray}
\label{fourier}
A^{\mu}(x,y)&=&\frac{\sqrt{2}}{\sqrt{2\pi
R}}A^{\mu}_{(0)}(x)+\frac{2}{\sqrt{2\pi
R}}\sum^{\infty}_{n=1}A^{\mu}_{(n)}(x)\cos\frac{ny}{R},~~~~
A^5(x,y) = \frac{2}{\sqrt{2\pi
R}}\sum^{\infty}_{n=1}A^5_{(n)}(x)\sin\frac{ny}{R}, \nonumber\\
\phi^+(x,y)&=&\frac{\sqrt{2}}{\sqrt{2\pi
R}}\phi^+_{(0)}(x)+\frac{2}{\sqrt{2\pi
R}}\sum^{\infty}_{n=1}\phi^+_{(n)}(x)\cos\frac{ny}{R},~~~~
\phi^-(x,y) = \frac{2}{\sqrt{2\pi
R}}\sum^{\infty}_{n=1}\phi^-_{(n)}(x)\sin\frac{ny}{R}, \nonumber\\ 
\mathcal{Q}_{i}(x,y)&=&\frac{\sqrt{2}}{\sqrt{2\pi 
R}}\bigg[(u_{i},d_{i})_{L}(x)+\sqrt{2}\sum^{\infty}_{n=1}\Big[P_L
\mathcal{Q}^{(n)}_{iL}(x)\cos\frac{ny}{R}+P_R
\mathcal{Q}^{(n)}_{iR}(x)\sin\frac{ny}{R}\Big]\bigg], \\
\mathcal{U}_{i}(x,y)&=&\frac{\sqrt{2}}{\sqrt{2\pi
R}}\bigg[u_{iR}(x)+\sqrt{2}\sum^{\infty}_{n=1}\Big[P_R
\mathcal{U}^{(n)}_{iR}(x)\cos\frac{ny}{R}+P_L
\mathcal{U}^{(n)}_{iL}(x)\sin\frac{ny}{R}\Big]\bigg], \nonumber\\
\mathcal{D}_{i}(x,y)&=&\frac{\sqrt{2}}{\sqrt{2\pi
R}}\bigg[d_{iR}(x)+\sqrt{2}\sum^{\infty}_{n=1}\Big[P_R
\mathcal{D}^{(n)}_{iR}(x)\cos\frac{ny}{R}+P_L
\mathcal{D}^{(n)}_{iL}(x)\sin\frac{ny}{R}\Big]\bigg], \nonumber
\end{eqnarray}
where $i=1,2$ correspond to first two generations. Above, $x\equiv
x^{\mu}$ ($\mu$=0,1,2,3 denote the four noncompact space-time
coordinates), $y$ denotes the fifth (extra) compactified coordinate,
and $M=$0,1,2,3,5. The fields $\phi^{\pm}_{(n)}(x)$ are the 4d KK
scalar fields, $A^{\mu}_{(n)}(x)$ are the 4d KK gauge fields, and
$A^{5}_{(n)}(x)$ are the 4d KK scalar fields in the adjoint
representation of the gauge group. The field $A^5(x,y)$ depends on
sine of $y$ to ensure its absence on the brane ($y=0$). The fields
$\mathcal{Q}, \mathcal{U}$ and $\mathcal{D}$ describe the 5d states
whose zero modes are the 4d SM quarks. The KK expansins of the
weak-doublet and -singlet leptons $\mathcal{L}$ and $\mathcal{E}$ are
not shown for brevity.

3. The third generation of the fermions 
   are 4d fields localised at the
   orbifold fixed point ($y = 0$):  
\begin{eqnarray} 
\mathcal{Q}_{3}(x,y)=\frac{\sqrt{2}}{\sqrt{2\pi
R}}\bigg[(t,b)_{L}(x)\bigg], ~~~
\mathcal{U}_{3}(x,y)=\frac{\sqrt{2}}{\sqrt{2\pi
R}}\bigg[t_{R}(x)\bigg], ~~~
\mathcal{D}_{3}(x,y)=\frac{\sqrt{2}}{\sqrt{2\pi
R}}\bigg[b_{R}(x)\bigg],
\end{eqnarray}
and similarly for the leptons.   

4. The Higgs components $\chi^{\pm}_{(n)}$ and $\chi^{3}_{(n)}$ mix with the
adjoint scalar fields ($W^{\pm 5}_{(n)}$ and $Z^{5}_{(n)}$) to produce
Goldstone bosons ($G^{0}_{(n)},G^{\pm}_{(n)}$) and three additional physical
scalar modes ($a^{0}_{(n)},a^{\pm}_{(n)}$), where,
\begin{eqnarray} 
G^{0}_{(n)}&=&\frac{1}{M_{Z_{(n)}}}\Big[M_{Z}\chi^{3}_{(n)}-\frac{n}{R}
Z^{5}_{(n)}\Big],~~~~ a^{0}_{(n)} =
\frac{1}{M_{Z_{(n)}}}\Big[\frac{n}{R}\chi^{3}_{(n)}+M_{Z}
Z^{5}_{(n)}\Big], \nonumber\\
G^{\pm}_{(n)}&=&\frac{1}{M_{W_{(n)}}}\Big[M_{W}\chi^{\pm}_{(n)}-
\frac{n}{R}
W^{\pm 5}_{(n)}\Big],~~~~ a^{\pm}_{(n)} =
\frac{1}{M_{W_{(n)}}}\Big[\frac{n}{R}\chi^{\pm}_{(n)}+ M_{W}W^{\pm
5}_{(n)}\Big],
\end{eqnarray} 
where $M_{n}^{2}=M^{2}+n^{2}/R^{2}$, $M$ being the zero mode $W$ or
$Z$ mass ($M_{W(Z)}$), and $M_n$ corresponds to a generic $n$th mode
mass.  Clearly, at the brane, only the $\chi_{(n)}$ states are
nonvanishing.

5. We identify two cases which require separate treatments.  Case (a):
We restrict our discussion to the first two generation of
fermions. Then all interactions and Feynman rules are exactly like in
UED discussed in detail in \cite{acd} and \cite{buras1}.  The KK
number is conserved at all vertices.  We, however, note that the
mixing of KK fermions within the same generation is controlled by an
angle $\alpha_n$, given by $\tan 2\alpha_{n}=m_{0}/M_{n}$. Since the
zero mode masses ($m_0$) of the first two generation of fermions are
negligible, we take this mixing to be vanishing.  Case (b): At least
one third generation fermion is involved. The lagragian contains
$\big[\delta (y)+ \delta (y-\pi R)\big]$ to ensure the localization of
this interaction {\em only} on the brane. There is no KK number
conservation as a result of breakdown of translational
invariance. Unlike in the UED case or case (a), the operator structure
of gauge boson - fermion interaction in case (b) is SM like.

\section{UV divergences for different processes in different models}
\subsection{\boldmath{$Z\bar bd$} vertex}
The effective $Z\bar bd$ vertex is constructed from triangle and self energy
diagrams in which the up-type quarks and $W$ boson circulate inside the
loops. For clarity, we use $W$ to denote the transverse part of the $W^\pm$
boson, while $\phi$ refers to its longitudinal component and the additional
scalars ($\phi \equiv G^\pm, a^\pm$). By $W$ (or $\phi$) mediated graphs we
mean loops with internal $W_{(n)}$ (or $\phi_{(n)}$) KK bosons. We assign an
index $i$ to indicate the three generations of up-type quarks inside the loop,
and use $\a$ to label the different diagrams for each $i$.  We employ a
notation $d^{W(\phi)}_\a$ which captures the relevant tree level couplings of
the $Z$ boson for the diagram labeled `$\a$' (i.e., $Zii$, or
$ZW(\phi)W(\phi)$ for triangles, and self energies on the $d$ and $b$ legs)
and the associated loop factors. The relevant Cabibbo-Kobayashi-Maskawa (CKM)
elements appearing in the loop vertices are separately denoted by $\xi_i
\equiv V_{id}V^*_{ib}$.

When we calculate the amplitudes, we integrate over the loop momentum ($k$)
and sum over the KK modes ($k_5 = n/R$). For individual loop graphs we
encounter two kinds of cutoffs: the 4d momentum cutoff denoted by $\Lambda$
(scaled to be a dimensionless number) and the KK momentum cutoff by $n_s$, and
we expect $\Lambda \sim n_s$.  Operationally, we perform the 4d integration
first and then do the KK summation\footnote{We stress that the operational
ordering of performing the 4d loop momentum integration first and then doing
the KK summation is technically more correct than the other way round. As we
stressed the 4d momentum cutoff $\Lambda$ should be of the order of the KK
momentum cutoff $n_s$. If we do the infinite KK summation first as in
Eq.~(\ref{kkinfinity}) and then perform the 4d momentum integration we include
contributions from scales above $\sim \Lambda$ which we have observed lead to
physicswise meaningless results. A similar conclusion has also been drawn in
\cite{nilles} in a different context.}.

The $W$ and $\phi$ mediated loop amplitudes have the following structures:
\begin{eqnarray} 
\label{geneq}
A_W = \sum_n A_{W_{(n)}}& = 
& \sum_n \sum_i \sum_\a \xi_i d^W_\a \left[\ln \Lambda 
                           + f_{i\a} (x_i, n)\right], \nonumber \\ 
A_\phi = \sum_n A_{\phi_{(n)}}& = 
& \sum_n \sum_i \sum_\a \xi_i x_i d^\phi_\a \left[\ln \Lambda 
                           + g_{i\a} (x_i, n)\right], 
\end{eqnarray} 
where $x_i \equiv m^2_{0i}/M_W^2$, $f_{i\a}$ and $g_{i\a}$ are finite pieces
obtained from the individual KK modes for the $W$ and $\phi$ mediated graphs
respectively. It is important to observe at this point that the relation
$\sum_i \xi_i = 0$ ensures through Glashow-Iliopoulos-Maiani (GIM) mechanism
the cancellation of the $\ln \Lambda$ dependence in the $W$ mediated graphs,
while the arrangement between wave function renormalizations and vertex
corrections leading to $\sum_\a d^\phi_\a = 0$ guarantees the absence of the
net $\ln \Lambda$ dependence in $\phi$ mediated loops.  We emphasize that
Eq.~(\ref{geneq}) is a kind of master equation which can cover all the three
extra dimensional scenarios under consideration.

{\bf 1. SM:}~ The Eq.~(\ref{geneq}) for $n = 0$ describes the SM situation.
As stated above, the GIM mechanism ($\sum_i \xi_i = 0$) and the relationship
between vertex corrections and wave function renormalizations ($\sum_\a
d^\phi_\a = 0$) together ensure the finiteness of the effective $Z\bar bd$
vertex.

{\bf 2. UED:}~ As mentioned above, the $\ln \Lambda$ dependence cancels out
mode by mode, so what remains to be seen is whether the KK sum over the finite
pieces from individual modes yields a finite or a divergent result. In this 
scenario
\begin{eqnarray}
\label{uedeq}
A_{W_{(n)}} \sim 1/n^2,  ~~~ 
A_{\phi_{(n)}} \sim  1/n^2,  
\end{eqnarray} 
in the large $n$ limit, and hence the KK summation over these individually
finite pieces also yields a finite effective $Z\bar bd$ vertex.

{\bf 3. NUED:}~ Since the external legs are all zero mode states, there is
still a single KK index running inside the loop.  In this case the $n$
dependence arises only from the bosonic excitations. Again, the amplitudes in
the large $n$ limit go like $1/n^2$.  Here also the effective $Z\bar bd$
vertex is finite.

{\bf 4. Split:}~ We recall that the first two generation of fermions have KK
excitations, while the third generation is brane localized. Still, an
inspection on the diagrams reveals that there is only a single KK index
running in the loop. The net $\ln \Lambda$ dependence cancels exactly for the
same reason cited in the context of the master equation (\ref{geneq}). Now we
ask the question whether KK summing over the finite parts of the individual
modes yields a divergent or a finite result. For that we again give a look at
the master equation. The large $n$ behaviour of the $W$ mediated graphs is
different now from the UED case, while in the same limit the $\phi$ mediated
loops behave as in UED. More specifically,
\begin{eqnarray}
\label{spliteq}
A_{W_{(n)}} \sim n^2/n^2, ~~~ 
A_{\phi_{(n)}} \sim  1/n^2,  
\end{eqnarray} 
where in the first case the appearance of $n^2$ in the numerator is a
consequence of the split nature, i.e. a relative localization of different
generation of fermions. The net divergence therefore appears only from this
$W$ mediated part after the KK summation, and it is not difficult to see that
it is a linear divergence. Taking $R^{-1} \sim 1 ~{\rm TeV} ~\gg m_t$, the
effective $Z\bar bd$ vertex looks like ($n_s \sim \Lambda$)
\begin{eqnarray}
\label{effzbdv}
\Gamma_{\mu}^{Zbd} = \frac{ig}{\cos\theta_{W}}\Big(\frac{g^2}{16\pi^{2}}
F_{S}\xi_{t}\Big)\gamma_{\mu}P_{L},~~~{\rm with}~~
F_{S} \simeq \left(\frac{3}{4} - {\frac{5}{6}} \sin^2\theta_W\right) \Lambda.
\end{eqnarray}

\subsection{\boldmath{$B-\bar B$} mixing}
{\bf 1. SM:}~  The relevant box diagrams are all finite. 

{\bf 2. UED:}~ Since KK number is conserved, there is only a single KK
number in the loop, hence a single KK summation. After one performs
the summation, it is not difficult to see from power counting that
each such box is finite. The bottom line is that the finiteness owes
to the single summation.  

{\bf 3. NUED:}~ There are two independent KK indices attached to the two
internal bosons. Their propagators can be summed independently, and thus each
such box (be it $W$ mediated or $\phi$ mediated) is log divergent. The
divergence from the ones involving two internal KK $W$'s sums up to zero on
account of GIM mechanism. But the divergences from the boxes having two KK
$\phi$'s just add up (the ones having one $W$ and one $\phi$ are finite
anyway). The dominant contribution goes as $\xi_t^2 x_t^2 \ln\Lambda$. The
double KK summation involved is the deciding factor behind this divergence.

{\bf 4. Split:}~ A point to observe is that two of the four vertices in the
box which connect to the external $b$ quarks necessarily violate the KK number
since the third generation quark is brane localized. The remaining two
vertices may or may not violate KK number (depending on whether the internal
fermion is $t$ quark or $u,c$ quarks). The result is that any such box
involves two independent KK summations, and hence each of them turns out to be
log divergent. The root of the net divergence not only lies in the $\phi$
mediated boxes, the $W$-boxes also sum up to a net divergence due to
incomplete GIM cancellation owing to the placement of different fermion
generations in different locations.

As an illustrative example, we present the functional dependence of the KK
modes in the amplitude (after the 4d loop momentum integration) when there are
two top quarks in the internal lines of a box ($y_n \equiv 1 + n^2/(R^2
M_W^2)$):
\begin{eqnarray} 
 I (n,m) & = & \frac{f(x_t, y_n) - f(x_t, y_m)}{y_n - y_m}, ~~~{\rm where} \\
 f(x_t, y_n) & = & \frac{x_t(0.5~x_t- y_n) \ln x_t+ 0.5~y_n^2 
\ln y_n - 0.75~y_n^2 - x_t(0.25~x_t - y_n)}{2(x_t-y_n)^2}.
\end{eqnarray}
Summing over $n$ and $m$ yields $I \sim \ln n_s \sim \ln \Lambda$. Other
combinations of the internal quark lines also lead to log divergence. The net
divergence structure expectedly reads $\sim \xi_{t}^{2} \ln\Lambda$ for the
$W$ mediated boxes and $\sim \xi_{t}^{2} x_t^2 \ln\Lambda$ for the $\phi$
mediated boxes.  We do not display here the exact coefficients, they are the
results of the above and other more complicated summations and intergrations.
On the other hand, the boxes containing one $W_{(n)}$ and one $\phi_{(m)}$ are
finite after KK summation.

\subsection{The \boldmath{$\rho$} parameter}
{\bf 1. SM:}~ Each of the $W$ and $Z$ self energy diagrams with fermion loops
is quadratically divergent, and each loop with internal bosons is log
divergent.  But the net contribution is UV finite having the well-known
expression which is approximately
\begin{equation}
(\Delta \rho)_{\rm SM} \sim 
\frac{\alpha}{\pi}\left[\frac{m_t^2}{M_Z^2} -\ln
\frac{M_h}{M_Z}\right].
\end{equation}

{\bf 2. UED:}~ The contributions coming from higher KK modes decouple
as their inverse square masses. The net contribution from each KK mode
is finite. It is interesting to note that for a given KK mode, the
contribution of $M_h$ to $\Delta \rho$ appears quadratically as
opposed to its logarithmic dependence in the SM contribution. The
contribution from the $n$th KK mode approximately reads \cite{acd}
\begin{equation}
\label{rhoued}
(\Delta \rho)_n \sim \frac{\alpha}{\pi}\left[\frac{m_t^4}{M_Z^2 M_n^2}
- \frac{M_h^2}{7M_n^2} - \frac{M_W^2}{M_n^2}\right].
\end{equation}  
Clearly, the KK summation leads to a finite result for $\Delta \rho$.

{\bf 3. NUED:}~ Since the fermions are all in the brane, new
contributions would come only from the KK bosons. The net contribution
is exactly the same as Eq.~(\ref{rhoued}) but without the first term,
i.e. without the fermionic part.

{\bf 4. Split:}~ The bosonic (gauge boson and Higgs) contribution would
expectedly be the same as in UED (the second and third terms in
Eq.~(\ref{rhoued})), and hence KK summation over the bosonic excitations leads
to a finite result. But the fermionic contribution is tricky, just because
different fermions are located at different places. First, it is important to
recall that the $Z$ boson self energy diagrams receive nonzero contribution
purely from the {\em axial} part of the fermionic coupling. The
$Z_{\mu}^{(0)}\bar\mathcal{U}_{i}^{(n)}\mathcal{U}_{i}^{(n)}$ and
$Z_{\mu}^{(0)}\bar\mathcal{Q}_{i}^{(n)}\mathcal{Q}_{i}^{(n)}$ couplings being
purely vectorial \cite{acd} do not contribute to the $Z$ boson self energy
diagrams. The $Z_{\mu}^{(0)}\bar\mathcal{U}_{i}^{(n)}\mathcal{Q}_{i}^{(n)}$
($i=1,2$) couplings are purely axial but still do not contribute to the extent
the mixing angle ($\alpha_{n}$) can be ignored for the first two generation of
fermions.  In the same limit, the $W$ self energy graphs also receive
vanishing contribution from the first two generation of fermions.  Only those
$W$ boson self energy diagrams in which one quark is a localized state,
i.e. $t(b)$, and the other is a bulk state, i.e.  $d$ or $s$ ($u$ or $c$),
remain {\em unmatched} in the sense that there are no corresponding $Z$ self
energy diagrams which could have possibly cancelled their divergences. Since
each diagram is divergent, these {\em unmatched} divergences from the $W$ self
energy graphs survive in the absense of their $Z$ counterpart. The net
divergence to $\Delta \rho$ arising from the surviving diagrams after the KK
summation is observed to go like $\Lambda^3$ {\em modulo} the suppression
factor $(1 - |V_{tb}|^2)$. Since in the SM each {\em individual} fermionic
graph has a quadratic divergence, a single KK summation enhances the degree of
divergence by one order.  The $\rho$ parameter thus offers the most serious
constraint to the split scenario.

\section{A summary of UV divergences and a possible remedy}
Our primary aim in this paper has been to study the UV cutoff
dependences in the split scenario and probe the root cause behind
them. In the process, we compare and contrast three scenarios: split,
UED and NUED, to make a judgement of their relative effectiveness.
The following points are worth noting.

1. If one takes more than one extra dimension, all such models give
divergent results at any order (even the tree amplitude diverges if
one can perform a summation over the KK propagator).
 
2. Suppose we restrict to one extra dimension only, and remain within
one loop. Then in the UED scenario all results are finite, while in
the NUED picture, quite a few processes are UV cutoff sensitive. The
core issue is whether KK number is conserved or not, which relates the
origin of divergence to the occurence of more than one KK summation
within a one loop integral. In the split scenario, the UV divergences
are more severe in some cases (the contribution to $\Delta \rho$ being
a glaring example). In the latter case, the appearance of divergence
is caused not only due to the occurence of more than one KK summation
in a one loop integral, but also as a result of incomplete GIM
cancellation since some of the fermions have KK excitations while some
are brane localized.

3. If one goes beyond one loop, then even insisting on one extra
dimension only, {\em all} such models will become UV sensitive with
varying degree of cutoff dependence.
 
A comparison of UV sensitivities of different models for different
processes has been summarised in Table 1.
\begin{table}
\begin{center}
\begin{tabular}{|c|c|c|c|}
\hline
Processes & UED & NUED & Split  \\                   
\hline
$Z\bar bd$ vertex & Finite & Finite & $\sim \Lambda$    \\ 
$B-\bar B$ mixing & Finite & $\sim \ln\Lambda$ & $\sim \ln\Lambda$  \\ 
$(\Delta\rho)_{\rm KK~fermion}$ & Finite & No Contribution & $\sim
\Lambda^{3}$ \\ \hline
\end{tabular}
\caption[]{{\sf Ultraviolet divergences in different cases, with one
extra dimension}.}
\end{center}
\end{table}

{\bf A possible remedy}: 

1. A presciption for a possible cure from the occurence of such divergence may
be advanced by supplementing these models with some ideas of brane
fluctuations advocated in \cite{abq}.  A bold, and rather far-fetched,
assumption will be to attribute some dynamics at the orbifold fixed points
which may render the couplings at the loop vertices to be KK label ($n$)
dependent. If in such a situation a typical coupling $g$ at a loop vertex is
replaced by $g(n)$, given by $g(n) \sim g a(n)~{\rm exp} (-cn^2/M_S^2R^2)$,
then the exponential suppression will lead to a well-behaved intergral at the
UV end.  Whether or not it can actually be realised in a concrete scenario is
beyond the scope of this paper.

2. Very recently, an idea which goes by the name of `minimal length scenario'
has emerged which does not admit a length scale smaller than the string length
($l_p \sim \hslash/M_S$) \cite{hossen1}. Be that as it may, even though the
momentum ($p$) can go to infinity, the wave vector ($k$) is restricted from
above by the requirement that the Compton wavelength ($2\pi/k$) cannot be
smaller than $l_p$. This means that the usual relation $p = \hslash k$ has to
be replaced, and a simple but quite {\em ad-hoc} ansatz would be to invoke
$l_p k(p) = {\tanh}(p/M_S)$ and $l_p \omega (E) = {\tanh}(E/M_S)$.  Admitting
this relation also for the compactified direction means an insertion of a
factor $(\partial k/\partial p) = {\rm sech}^2 (p/M_S)$ inside a KK summation
where $p$ is equivalent to the KK mass $M_n$. Since ${\rm sech}(x) \sim {\rm
  exp}(-x)$ for large $x$, one obtains an exponential suppression of the
effective coupling (or, form factor) for higher KK states\footnote{There is a
  slight difference between the string- and minimal length- form
  factors. While the string form factors are rather of exp ($-p^2$) type, the
  minimal length principle, as advocated in \cite{hossen1}, generates a form
  factor which has a linear dependence on ($-|p|$) in the exponential.}.  We
must confess that from a pure field theoretic perspective all we did is a
simple trading of the cutoff parameter $\Lambda$ in favour of an {\em ad-hoc}
form-factor which contains a minimal length $l_p$ in the way shown. One should
note however that if we agree to go by the above hypothesis then the
appearance of an exponentially suppressed form-factor ensures a smooth loop
integration up to infinity at all orders.

\section*{Acknowledgements}
We thank E. Dudas and A. Raychaudhuri for reading the manuscript and
suggesting improvements. We also thank S. SenGupta for discussions and for
bringing \cite{nilles} to our attention. G.B. thanks K. Sridhar for bringing
the first paper in \cite{hossen1} to his notice and acknowledges discussions
on minimal length issues with E. Dudas, P. Mathews and K. Sridhar. G.B. also
thanks A. Raychaudhuri for discussions on issues of fermion flavor in extra
dimensional theories. G.B.'s research has been supported, in part, by the DST,
India, project number SP/S2/K-10/2001.

\end{document}